\documentclass[useAMS,usenatbib]{mn2e}

\def\Msun{\hbox{$\rm\thinspace M_{\odot}$}}
\def\Rsun{\hbox{$\rm\thinspace R_{\odot}$}}
\def\Re{\hbox{$\rm\thinspace R_{\oplus}$}}
\def\Mjup{\hbox{$\rm\thinspace M_{\textrm{\tiny jup}}$}}
\def\Rjup{\hbox{$\rm\thinspace R_{\textrm{\tiny jup}}$}}

\def\Rwd{\hbox{$\rm\thinspace R_{\textrm{\tiny WD}}$}}

\def\ga{\mathrel{\hbox{\rlap{\hbox{\lower4pt\hbox{$\sim$}}}{\raise2pt\hbox{$>$}}}}}
\def\la{\mathrel{\hbox{\rlap{\hbox{\lower4pt\hbox{$\sim$}}}{\raise2pt\hbox{$<$}}}}}

\usepackage{color}
\usepackage{epsfig}
\usepackage{lscape}
\usepackage{supertabular}
\usepackage{graphicx}

\title[Sub-stellar and planetary companions to WDs]{Detection limits for close eclipsing and transiting
  sub-stellar and planetary companions to white dwarfs in the WASP
  survey}

\author[F. Faedi et al.]
  {F. ~Faedi,$^{1,2}$\thanks{E-mail:f.faedi@qub.ac.uk}
  R. G.~West,$^1$ M. R.~Burleigh,$^1$ M. R.~Goad,$^1$
  L.~Hebb,$^{3,4}$ 
\\
  $^1$Department of Physics and Astronomy, University of Leicester, University Road, Leicester, LE1 7RH, U.K.\\
  $^2$Astrophysics Research Centre, School of Mathematics and Physics, Queens University, University Road, Belfast, BT7 1NN, U.K.\\
  $^3${School of Physics and Astronomy, University of St. Andrews, North Haugh, Fife, KY16 9SS, U.K.}\\
  $^4${Department of Physics and Astronomy, Vanderbilt University, Nashville, TN 37235, U.S.A.}}
\date{Accepted 2010 August 5. Received 2010 August 3; in original form 2010 July 13}

\pagerange{\pageref{firstpage}--\pageref{lastpage}} \pubyear{2010}

\def\LaTeX{L\kern-.36em\raise.3ex\hbox{a}\kern-.15em
    T\kern-.1667em\lower.7ex\hbox{E}\kern-.125emX}

\begin{document}

\label{firstpage}

\maketitle

\begin{abstract}
{We have performed extensive simulations to explore the possibility of detecting eclipses and  
transits of close, sub-stellar and planetary companions to white dwarfs in WASP light curves. 
Our simulations cover companions $\sim0.3\Re<{\rm R}_{pl}<12\Re$ 
and orbital periods $2{\rm h}<P<15{\rm d}$, equivalent to orbital
radii $0.003{\rm AU} < a < 0.1{\rm AU}$. For Gaussian random noise
WASP is sensitive to transits by companions as small as the Moon
orbiting a $\textrm{V}\simeq$12 white dwarf. For fainter white dwarfs WASP is sensitive 
to increasingly larger radius bodies. However, in the presence of correlated
  noise structure in the light curves the sensitivity drops, although 
  Earth-sized companions remain detectable in principle even in
  low S/N data. Mars-sized, and even Mercury-sized bodies yield
  reasonable detection rates in high-quality light curves with little
  residual noise. We searched for eclipses and transit signals
  in long-term light curves of a sample of 194 white dwarfs resulting
  from a cross-correlation of the McCook $\&$ Sion catalogue and the
  WASP archive. No evidence for eclipsing or transiting sub-stellar and
  planetary companions was found. We used this non-detection and
  results from our simulations to place tentative upper limits to the
  frequency of such objects in close orbits at white dwarfs. While
  only weak limits can be placed on the likely frequency of
  Earth-sized or smaller companions, brown dwarfs and gas giants
  (radius $\approx \Rjup$) with periods $<0.1-0.2$~days must certainly
  be rare ($<10\%$). More stringent constraints likely requires
  significantly larger white dwarf samples, higher observing cadence
  and continuous coverage. The short duration of eclipses and transits
  of white dwarfs compared to the cadence of WASP observations appears
  to be one of the main factors limiting the detection rate in a
  survey optimised for planetary transits of main sequence stars.}
\end{abstract}

\begin{keywords}
method: data analysis - occultations - stars: planetary systems - stars: white dwarfs

\end{keywords}

\section{Introduction}

In recent years we have witnessed considerable progress in the search
for extra-solar planets. Since the first detection of a `Hot
Jupiter' around the main-sequence star 51 Peg (\citealt{Mayor95}),
the number of extra-solar planets has rapidly risen, and currently
approaching {500}. Most of these discoveries are the result of radial
velocity (RV) searches. More recently, an increasing number of
extra-solar planets ($>80$) have been detected by dedicated planetary
transit surveys including HATnet (\citealt{Bakos04}), TrES (e.g.
\citealt{Brown00}; \citealt{Dunham04}; \citealt{Alonso04}), OGLE
(\citealt{Udalski02}, 2003), XO (\citealt{McCullough05}), and WASP,
the UK Wide-Angle Search for Planets (\citealt{Pollacco06}).

Planet detection via the transit technique involves searching for
periodic dips in stellar light curves as a planet occludes a small
fraction of the visible disc of the host star once per orbit. Only
planets with their orbital planes aligned within a few degrees to the
line of sight will exhibit a transit, the probability of such an
alignment being around 10\% for typical `hot Jupiter' systems. This
introduces a constraint on the number of observable systems and
explains the relatively low number of transiting planets when compared
to radial velocity studies. Importantly, when combined with RV
measurements, planetary transits offer the unique possibility of
deriving both the planet mass and radius, since for these systems the
inclination $i$ is well-known \citep{Sackett99}. For a given planetary
radius, the transit depth is directly proportional to (R$_{\rm
  p}$/R$_{\rm *})^{2}$, where R$_{\rm p}$, and R$_{\rm *}$ are the
planetary and stellar radii respectively. Therefore, planets orbiting
solar-type stars have extremely shallow eclipses, blocking $\sim$1\%
of the light for a giant planet and $\sim$0.01\% of the light for an
Earth-sized planet. Current ground-based wide-field surveys can
achieve the necessary photometric accuracy of better than 1$\%$, only
for the brightest stars ($\textrm{V}\sim9$--12 in the case of WASP), so the
bulk of the planets discovered by transit surveys around main-sequence
stars have radii in the range $R_p\sim 0.9$--1.8\Rjup. To date the
smallest extra-solar planet detected in a ground-based transit survey
is HAT-P-11b, a Neptune-size planet ($R_p=0.452$\Rjup) transiting a K
dwarf star (\citealt{Bakos10}).

A major advantage over main sequence primaries is offered by white
dwarf stars. White dwarfs (WDs) are compact degenerate objects with
\Rwd$\sim 1$\Re (Earth radius), and represent the final stage of
evolution of main$-$sequence stars with masses $\leq$ 8\Msun (i.e.
$\sim 97\%$ of all stars in our galaxy). Any sub-stellar or gas giant
companion orbiting the star, will completely eclipse it, while bodies
as small as the Moon will display relatively large transit depths
($\sim$ 3$\%$), with the only caveat being that it remains unclear as
to whether any such systems survive beyond the latter stages of
stellar evolution. The strong gain in the planet$-$to$-$star relative
dimensions opens up the possibility of detecting low$-$mass
sub$-$stellar and in particular terrestrial objects in orbit around
WDs. In Sections~1.1, 1.2 we briefly discuss theoretical
studies concerned with the likelihood of sub-stellar and planetary
survival to stellar evolution.

\subsection{Sub-stellar companions to WDs}

Observationally, sub-stellar companions to WDs are found to
be rare. Using the 2MASS survey, \citet{Farihi05} estimated that
$<0.5\%$ of WDs have L dwarf companions. More recently,
excess near-infrared emission from WDs in the UKIDSS survey
(Steele et al. in preparation) tentatively suggests the fraction of
unresolved brown dwarf companions (including T dwarfs) may be slightly
higher, between $1-2\%$. However, at the time of writing only three
wide WD+BD systems have been spectroscopically confirmed, GD\,165
\citep{Becklin88}, PHL5038 \citep{Steele09}, and LSPM~$1459+0857$\,AB
\citep{DayJones10} and two detached, non-eclipsing, short-period WD+BD
systems are currently known, WD$0137-349$ (\citealt{Maxted06},
\citealt{wd0137b}, $P \approx116$m), and GD1400 (\citealt{Farihi04},
\citealt{Dobbie05}, Burleigh et al., in preparation, $P \approx9.9$h).
GD1400B and WD$0137-349$B are the only two sub-stellar companions
known to have survived the common envelope (CE) phase of stellar
evolution, with WD$0137-349$B currently the lowest mass ($\sim50
\Mjup$) object known to have done so.

Although infrared sky surveys such as UKIDSS, VISTA and WISE, and
observatories such as Spitzer hope to reveal many more such binaries,
they remain difficult to identify either as infra-red excesses or
through radial velocity measurements. The detection of more close
systems will allow us to place observational upper limits on the mass
of sub-stellar companions that can survive CE evolution. Furthermore,
examples of eclipsing WD+BD binaries will be important for exploring
the WD and substellar mass-radius relations
(e.g.~\citealt{NNSer}).

In addition the detection of a significant number of eclipsing WD+BD
binary systems might help uncover the hypothesised population of `old'
cataclysmic variables (CVs) in which the current accretion rate is
extremely low and the companion has been reduced to substellar mass
(e.g. \citealt{Patterson98}; \citealt{Patterson05};
\citealt{Littlefair03}). While these systems elude direct detection as
X-ray sources and remain difficult to identify in optical and
infra-red surveys, it is possible to measure the mass and the radius
of the donor in eclipsing CVs. \citet{Littlefair06} confirmed the
first such system through eclipse measurements, while
\citet{Littlefair07} showed that another eclipsing CV,
SDSS~J$150722.30+523039.8$, was formed directly from a detached WD/BD
binary. Old CVs are important for shedding light on models of close
binary evolution as well as for placing constraints on the period
distribution of cataclysmic variables; in particular, the period gap
and the period minimum (\citealt{King88}; \citealt{Parthasarathy07}).

\subsection{Can planets survive stellar evolution?}

Every star less massive than 8\Msun~($\sim$ 97$\%$ of all stars in our
galaxy) will end its life as a WD. Thus, it is natural to ask what
will be the fate of known extra-solar planetary systems? This question
also has particular interest for us, in that the Earth's survival to
the Sun's post-main sequence evolution is uncertain
(\citealt{Rasio96}; \citealt{Duncan98}; \citealt{Villaver07}). Several
theoretical studies discuss post-main sequence evolution of planetary
systems and show that planetary survival is not beyond possibility
(\citealt{Duncan98}; \citealt{Debes02}; \citealt{Burleigh02}; and
\citealt{Villaver07}). Radial velocity observations of red giants
indicate that planets orbits beyond the radius of the star's envelope
can survive stellar evolution to that stage (see \citealt{Frink02};
\citealt{Hatzes05}, \citealt{Sato03}). However, direct imaging
searches at WDs have so far failed to detect any planetary mass
companions (e.g. \citealt{Hogan}). More recently, \citet{Silvotti07}
reported the detection of a $\sim$ 3\Mjup~planet orbiting an extreme
horizontal branch star. Furthermore, \citet{Mullally08} found
convincing evidence of a 2\Mjup~planet in a 4.5 year orbit around a
pulsating WD. The latter, if confirmed, will be the first planet
detected in orbit around a WD, and will show that planets can indeed
survive the death of their parent star.

The existence of short-period planetary companions to WDs may seem
less likely. Two scenarios may give rise to planets in short-period
orbits around WDs: \newline 1) planets undergo CE evolution and
survive their parent stars' evolution to a WD, or 2) their orbits are
significantly changed by a process occurring at the end of the
asymptotic giant branch (AGB) phase of stellar evolution.

\citet{Villaver07} investigated the fate of a planet engulfed by the
envelope of an AGB star and suggested that planets in orbit within the
reach of the AGB envelope will either totally evaporate or in rare
cases, a more massive body may accrete mass and become a close
companion to the star. In this scenario only the massive companions
(e.g. brown dwarfs like GD1400 and WD$0137-349$B) are likely to
survive the red giant and the AGB phases of stellar evolution.
However, estimates of the minimum substellar mass necessary for
survival are highly uncertain and depend on several factors, for
example, the efficiency of the envelope ejection (\citealt{Villaver07}
and references therein). None the less, it is unlikely that
terrestrial planets can survive engulfment and evaporation
\citep{GD356}.

Planets that escape engulfment by the red giant or asymptotic giant
and that are sufficiently far from the stellar surface that they do
not experience tidal drag, will have their orbital radii increased to
conserve angular momentum (as described by \citealt{Jeans24}).
\citet{Duncan98} investigated the stability of planetary systems
during post main-sequence evolution, and found that for WD progenitors
experiencing substantial mass loss during the AGB phase, planetary
orbits become unstable on timescales of $\leq$10$^{8}$~year.
\citet{Debes02} also studied the stability of planetary systems and
found that mass loss from the central star is sufficient to
destabilise planetary systems comprising two or more planets. For
unstable systems in which the orbits happen to cross, \citet{Debes02}
found that the most likely result was that one planet would be
scattered into an inner orbit, while the other would either be boosted
into a larger orbit, or ejected from the system altogether. This may
result in WD systems which have settled into a configuration
wherein planets are found at orbital radii which were originally
occupied by the (now evaporated) inner planets before the RGB phase of
stellar evolution.

The above scenario provides a plausible explanation for the recent
detection of silicate-rich dust discs around a growing number of WDs
at orbital radii up to $\sim$ 1{\Rsun} (e.g. \citealt{Jura03};
  \citealt{Reach05}; \citealt{Farihi08b};
  \citealt{Farihi09}).\citet{Jura03} suggests that the formation of
  dust discs around WDs is most probably due to the tidal disruption
  of an asteroid or larger body which has strayed too close to the
  parent star. Dynamical instabilities during the final stages of
  solar system evolution could have caused the rocky body to migrate
  inwards (as suggested by \citealt{Debes02}). If the body wanders too
  close to the Roche radius of the WD it will be completely
  destroyed, producing a debris disc reminiscent of Saturn's rings
  \citep{Jura03}. Recent studies of the dust disc around the WD
  GD\,362 (\citealt{Jura09}) suggest that the more likely
  scenario which simultaneously explains all of GD\,362's distinctive
  properties is that we are witnessing the consequences of the tidal
  destruction of a single body that was as massive as Callisto or
  Mars.

  Consideration of dynamical interactions and orbital stability
  indicates that while a terrestrial body may be perturbed from some
  wide orbit into an eccentric orbit that takes it within the Roche
  radius of a WD and hence be disrupted, it is highly unlikely for
  such an object to be captured into a stable, close orbit just beyond
  the Roche radius. None the less, one may speculate on the existence
  of `shepherd moons' accompanying the dust discs detected at WDs,
  similar to those at Saturn's rings. Alternatively, when
  close, double WDs are drawn together by gravitational radiation and
  merge, second generation terrestrial planets may form in remnant
  discs left by the tidal disruption of the lower mass degenerate
  (\citealt{Hansen02}, \citealt{LPW05}). Indeed, \citet{GD356}
  speculates that such an object may be closely orbiting the unusual
  magnetic WD GD\,356. Thus, the existence of asteroids, moons and
  rocky planets in close orbits to WDs may not be entirely
  unreasonable.

The detection of short-period substellar and planetary-mass
companions to WDs, will open an exciting chapter in the study
of extra-solar planet evolution, constraining theoretical models of
common envelope evolution and helping us to understand the ultimate
fate of hot Jupiter systems as well as the fate of our own solar
system in the post main-sequence phase. In this work we present the
results of a study designed to investigate the detection limits for
transiting sub$-$stellar and terrestrial companions in close orbits
around WDs \footnote{We note that \citet{DiStefano} discuss
  the possibility of discovering asteroids and moons in much wider
  orbits around WDs in Kepler data.}. In Section~2 we describe briefly
the WASP project and WASP telescopes, which provide the WD
light curves which form the basis of our transit search. In Section~3
we outline our Monte Carlo simulations, describing our detection
method, as well as characterising the type of systems we might hope to
detect. In Section~4 we discuss the results of our simulations and
provide transit recovery rates for simulated light curves comprising
random Gaussian noise (white noise) and correlated noise (red noise).
In Section~5 we present the results of a comprehensive transit search
in a sample of 194 WD light curves found in the WASP archive.
Finally, our conclusions are presented in Section~6.

\section{The WASP project}

The WASP (Wide-Angle Search for Planets) project, operates two robotic
telescopes, one located amongst the Isaac Newton Group of telescopes
(ING), in La Palma Spain, with a second instrument situated at the
South African Astronomical Observatory (SAAO). Each instrument
consists of eight f/1.8 Canon lenses each with an Andor CCD array of
2048$^{2}$ 13.5$\mu$m pixels, giving a field of view of 7.8 degrees
square for each camera. The observation strategy is to cyclically
raster the sky in a series of fields centred on the current local
sidereal time and separated by 1 hour in right ascension. Each
observation lasts for about 1 minute (30 seconds exposure, plus slew
and telescope settling time). This strategy yields well sampled
light curves with a typical cadence of about 8 minutes per field. WASP
provides good quality photometry with accuracy $\le 1\%$ in the
magnitude range $\textrm{V}\sim9$--12. The WASP telescopes and data analysis
strategies are described in detail in \citet{Pollacco06}.

\section{Detectability of eclipses and transits of white dwarfs}

To assess the chances of detecting eclipses and transits of white
dwarf host stars in ground-based wide-field surveys we performed an
extensive set of Monte Carlo simulations. The approach we adopted was
to create realistic synthetic light curves containing eclipse and
transit signatures of the expected depth and duration for a range of
companion size and orbital period, then to attempt to detect these
signatures using a standard transit-detection algorithm
(box-least-squares, BLS). By noting the rate at which the BLS search
recovered the transit at the correct period (or an integer multiple or
fraction) we were able to estimate the feasibility of detecting such
systems in an automated manner.

\subsection{Characteristics of the transit signal}

The probability ($p_{\rm tr}$) that a low mass star, brown dwarf or planet in a
circular orbit will transit or eclipse its host star is given by

\begin{equation}\label{eqprob}
p_{\rm tr}\simeq\left(\frac{4\pi^2}{GM_*}\right)^\frac{1}{3}\frac{R_p+R_*}{P^{2/3}}
\end{equation}

Assuming an orbital inclination $i=90\degr$ the depth ($\delta_{\rm tr}$) 
and duration ($D_{\rm tr}$) of such a transit are given by

\begin{equation}\label{eqdepth}
\delta_{\rm tr}=\frac{\Delta F}{F}=\cases{R_p^2/R_*^2,
  & for $R_p\leq R_*$\cr
1, & for $R_p>R_*$}
\end{equation}

\begin{equation}\label{eqduration}
  D_{\rm tr}=2\sqrt{\frac{a}{GM_*}}(R_p+R_*)
\end{equation}

For our simulations we have chosen the parameters of the host star to
represent a typical 1-\,Gyr-old carbon-core WD of mass
$M_*=0.6$\Msun\ and radius $R_*=0.013$\Rsun. We explored the
detectability of planetary transits across the two-dimensional
parameter space defined by the orbital period and the planet radius.
We considered orbital periods in the range $P\sim 2$ hours to 15 days
(equivalent to orbital distances of between $a\sim 0.003$ and
0.1\,AU). The lower-limit to the orbital period was chosen to yield an
orbital separation close to the Roche radius of the WD, the
upper-limit by a requirement that we have a reasonable chance of
detecting five or more transits in a typical 150 day observing season
of a WASP survey field.

\begin{figure} 
\centering
  \psfig{file=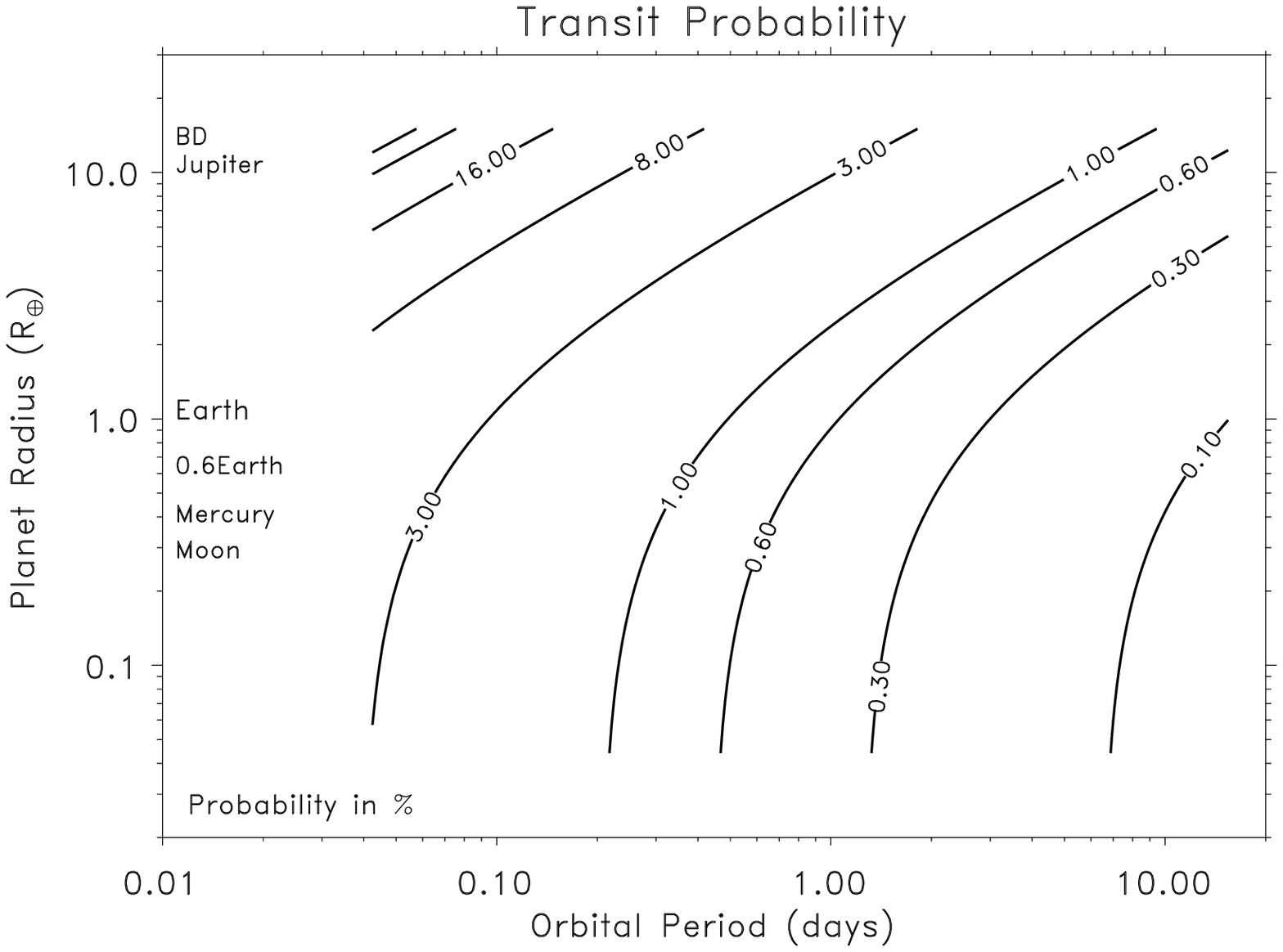,width=0.4\textwidth}
  \psfig{file=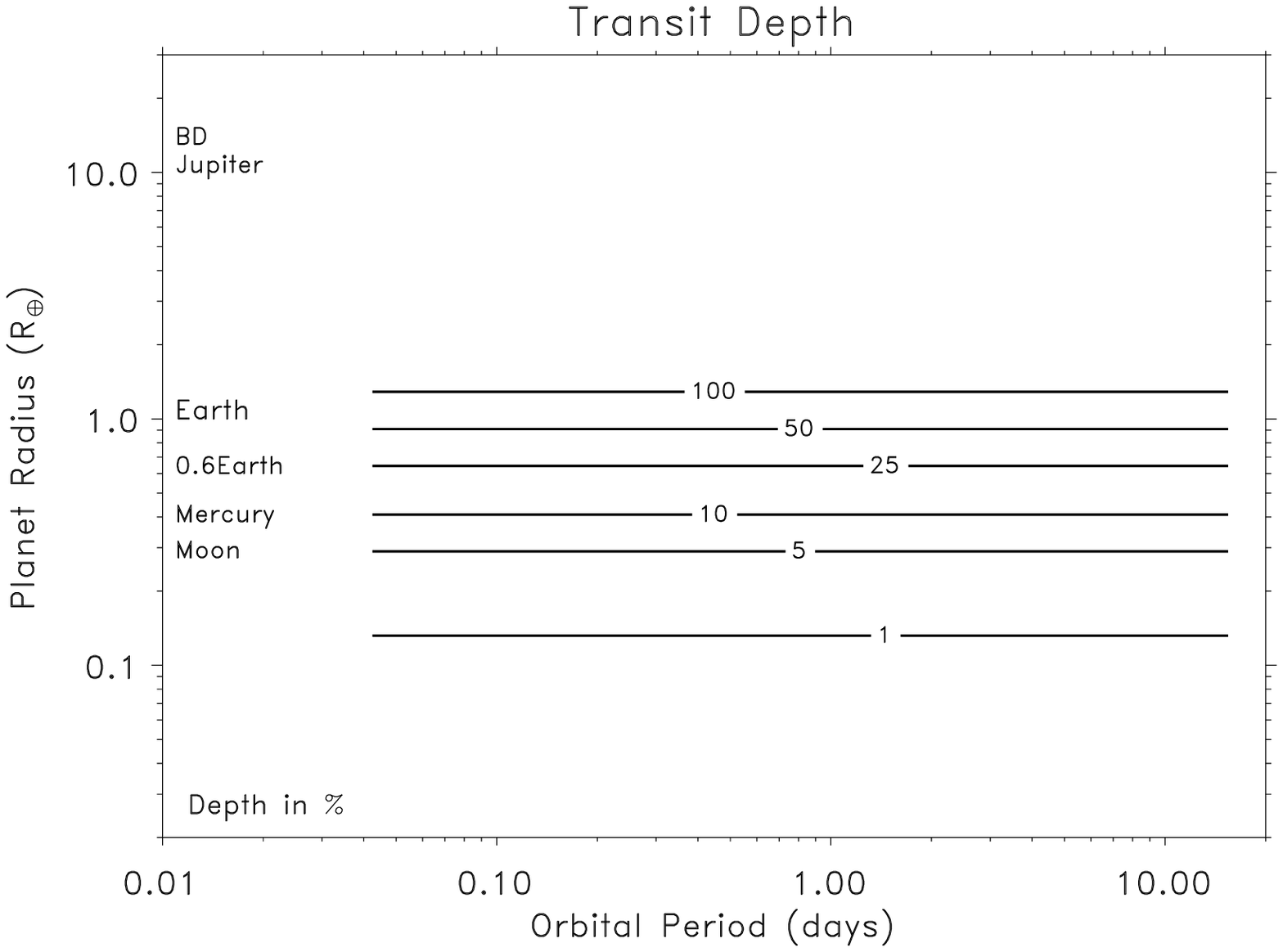,width=0.4\textwidth} \hfill
  \psfig{file=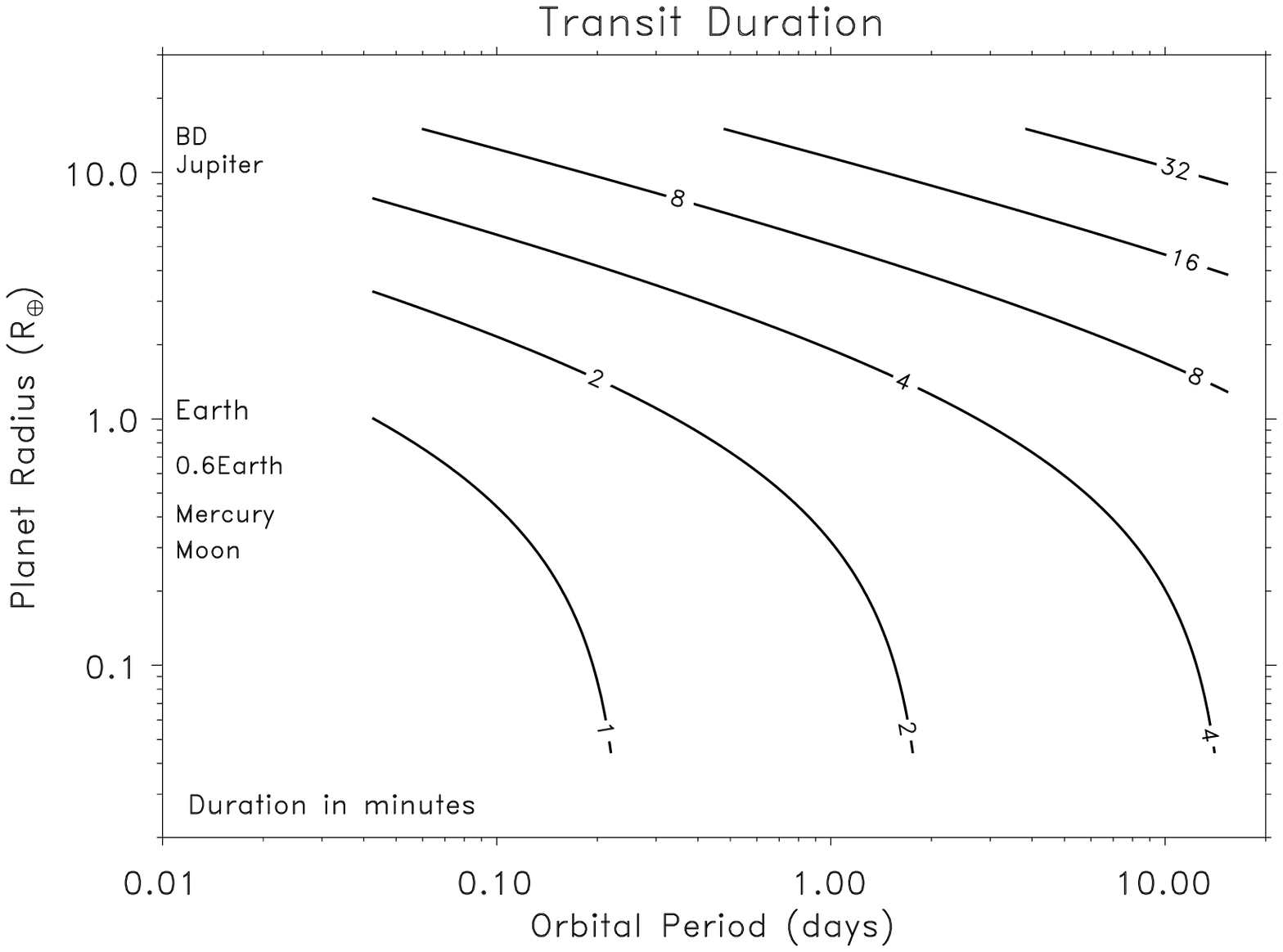,width=0.4\textwidth} \hfill
\caption{Contours of constant transit
    probability (top), depth (centre) and duration (bottom)  in the parameter
    space defined by orbital period and planetary radius. The transit 
    probability and depth are expressed in percentage values. The transit
    duration is expressed in minutes.}
  \label{paramspace} 
\end{figure}
\noindent

Figure~\ref{paramspace} shows the probability that a given system will
transit, and the depth and duration of such transits across this
parameter space. It is evident from this diagram that the signatures
of transits of WDs by typical planet-sized bodies will be
rather different than those seen for typical transiting hot Jupiters.
In particular the transit duration is much shorter for WDs
than for normal stars (from $\sim1$--30~min for companions with sizes
ranging from Moon-size to Jupiter-size, compared to 2--3 hours for a
typical hot Jupiter), and the transit depths are much larger (from
around 3\% for a Moon-sized to 100\% for any companion larger than the
Earth, compared to $\sim1$\% for a hot Jupiter).

\subsection{Generation of synthetic light curves}

The synthetic light curves were generated using the time sampling of a
typical WASP survey field. light curves were generated with
statistical signal-to-noise ratios representative of three magnitude
ranges ($\textrm{V}\simeq12$,~13 and~15) spanning the range of
brightness of WDs in the WASP survey. The corresponding
photometric accuracy of WASP over this range is $\sim$1\% to 10$\%$.
Because WASP data show residual covariant-noise structure due to
instrumental systematics we have tested the transit recovery rate in
the case of both uncorrelated ``white'' noise and correlated ``red''
noise. In the white-noise case we injected transit-like signatures
into otherwise non-variable light curves, adding a
Gaussian-distributed noise component of standard deviation $\sigma$.
We chose $\sigma$ to be representative of the mean photometric error
on the points obtained from a real WASP light curve for an object of
our chosen magnitude. The template WASP light curves therefore defined
the time-sampling and the average signal-to-noise, but the photometry
was otherwise entirely synthetic. In the red noise case we injected
fake transits into a set of unmodified WASP light curves obtained from
a densely sampled field observed during the 2004 season. Data from the
2004 season have been detrended and thoroughly searched for
transit-like events as described in \citet{Cameron06}. Moreover, we
cross-correlated stars in the WASP field with the publicly available
General Catalogue of Variable Stars (GCVS; \citealt{Samus04}), and we
evaluated the $RMS$ of each light curve which we used to identify and
remove variable objects after individual eye-balling. Finally, each
light curve contained around 4240 data points, acquired over 116
nights, and spread across a baseline of 128 nights, with photometric
accuracy raging from $\sim1\%$ to $\sim10\%$ for stars in the
magnitude range $12<V<15$ . For each light curve we used the WASP
pipeline fluxes and errors derived after de-trending by the SysRem
algorithm~\citep{Cameron06}.

Planet transit light curves of main-sequence stars show a
characteristic shape, with an ingress lasting several tens of minutes,
a flat bottom of 2--3 hours and an egress again lasting tens of
minutes. For the case of a WD host star considered here, the
ingress and egress duration is typically short compared to cadence of
the WASP survey (8--10 minutes). We therefore ignore the detailed
shape of the ingress and egress phases and modelled the transit
signatures as simple box-like profiles.

To cover the orbital period-planet radius parameter space we selected
seven trial periods spaced approximately logarithmically ($P=0.08$,
0.22, 0.87, 1.56, 3.57, 8.30 and 14.72 days), and five planet radii
$R_{\rm p}=10.0$, 1.0, 0.6, 0.34 and 0.27\,\Re. We modelled the set of
synthetic light curves by injecting fake transit signals into
phase-folded light curves at the trial period with a random transit
epoch $t_0$ in the range $0<t_0<P$. We computed the transit duration
according to Equation~\ref{eqduration}, and hence the width of the
transit in orbital phase $\phi_{\rm tr}=D_{\rm tr}/P$. For all data
points falling in the phase range $0\leq\phi_{i}\leq\phi_{\rm tr}$ we
then reduced the observed flux by a factor $\delta_{\rm tr}$. For each
combination of orbital period and planet radius we generated 100
synthetic light curves.

\begin{figure}
  \centering \psfig{file=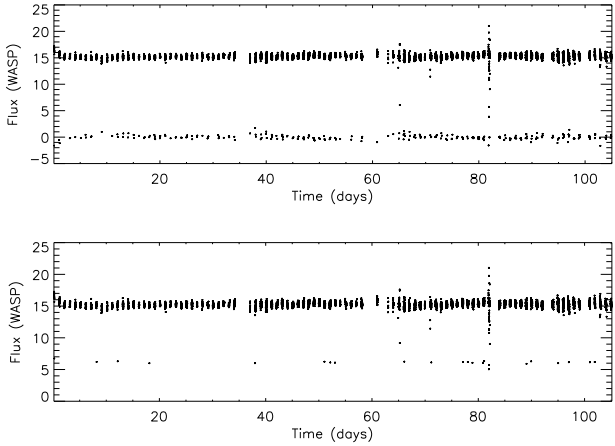,width=0.5\textwidth}
  \caption{Two example synthetic light curves. Top, an eclipsing brown dwarf in an orbit
    with a period of 2\,hr, around a WD. Bottom, a $1.2\Re$
    companion to a WD in 5\,hr orbit.}
\label{lightcurves}
\end{figure} 

Figure \ref{lightcurves} shows two examples of our simulated transit
light curves. The top panel shows the synthetic light curve of an
hypothetical eclipsing WD$+$BD binary system with an orbital period of
$P=116$~mins, similar to WD$0137-349$ (a non-eclipsing system,
\citealt{Maxted06}). The lower panel shows the simulated transit
light curve for a rocky body of radius $1.2\Re$ in a
5\,hr orbit.\\

\subsection{Detection algorithm}

To recover the transit signals from the synthetic light curves we used
an implementation of the box-least-squares (BLS) algorithm
\citep{Kovacs02} commonly used to detect transits of main sequence
stars. The BLS algorithm is most sensitive when the modelled box-width
closely matches the duration of the true transit signal. Thus, to
ensure that the BLS search was sensitive across the expected range of
transit durations, we chose to search a grid of box widths $W_{\rm
  b}=\{1, 2, 4, 8, 16, 32\}$ minutes, covering the range in transit
durations over most of our parameter space (Figure~\ref{paramspace}).
We defined the grid of trial periods sampled by BLS as follows:

\begin{equation}
F_{\rm max}=\frac{1}{{P}_{\rm min}} , \qquad F_{\rm min}=\frac{1}{{P}_{\rm max}}
\label{fgrid}
\end{equation}
\noindent
where $P_{\rm min}=2$\,hr and $P_{\rm max}=15$\,day. The frequency
interval was chosen such that the accumulated phase difference between
successive trial frequencies over the duration of the light curve
corresponds to the width of the shortest trial box duration at the
longest period searched. At each trial frequency we defined a set of
trial transit epochs at an interval $W_{\rm b}'$ chosen such that
$W_{\rm b}'\simeq W_{\rm b}$ for the shortest trial box duration,
adjusted to meet a constraint that the number of epochs $N_e=P/W_{\rm
  b}'$ be an integer.

Adopting the notation of \citet{Cameron06} we denote the set of
observations in the light curve $\tilde{x_i}$ with formal variances
$\sigma^2_i$ and additional variances $\sigma^2_{t(i)}$ computed by
{\sc SysRem} to account for transient systematic variations due to
patchy atmospheric extinction, for example. For each data point we
compute a weight

\[
w_i=\frac{1}{\sigma^2_i+\sigma^2_{t(i)}}
\]

then subtract the weighted mean of the observations

\[
\hat{x}=\frac{\sum_i\hat{x_i}w_i}{\sum_i w_i}
\]

to obtain $x_i=\tilde{x_i}-\hat{x}$. We then define

\[
t=\sum_i w_i
\]

summing across the whole dataset. At each trial period we fold the
light curve and accumulate into a set of bins $j$ of width $W_{\rm
  b}'$ the following quantities

\[
s_j=\sum_{i\in j}x_i w_i , \qquad r_j=\sum_{i\in j}w_i
\]

The fitted transit depth in the bin, and its associated variance are then

\[
\delta_j=\frac{s_j t}{r_j(t-r_j)} , \qquad {\rm Var}(\delta_j)=\frac{t}{r_j(t-r_j)}
\]

and the signal-to-noise ratio of a putative transit in the bin is:

\begin{equation}
\mathcal{S}_j=\frac{\delta_j}{\sqrt{{\rm Var}(\delta_j)}}
\label{sn_unmod}
\end{equation}

For each epoch bin we compute the signal-to-noise $\mathcal{S}_{j,b}$
for each of our trial box widths by co-adding values of $s_j$ and
$r_j$ in adjacent bins and find the maximum $\mathcal{S}_{j,max}$ over
all of the box widths. The maximum value of $\mathcal{S}_{j,max}$
across all of the epoch bins then represents the significance of the
detection of a transit at the given trial period. When computing this
latter maximum we ignored trial boxes which contained less than five
data points, or did not contain data points from at least five
distinct orbits.

\begin{figure} 
\centering
  \psfig{file=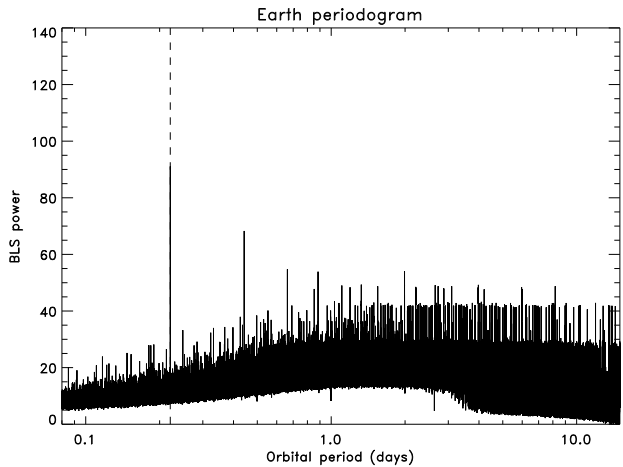,width=0.5\textwidth} 
\caption{BLS power spectrum for the transit signal of an Earth-sized body in
    an orbit with a 5 hour period. While the correct (inserted) period is recovered (as indicated by the dashed line)
    by the standard implementation of the BLS algorithm, strong
    aliasing and noise structure are present, particularly at long
    periods.}
\label{old_prdg}
\end{figure}

Figure~\ref{old_prdg} shows the periodogram computed in this fashion
for the sample light curve shown in the lower panel of
Figure~\ref{lightcurves}, a $1.2\Re$ radius body in a 5 hour orbit.
Although the correct simulated period shows up as the highest peak in
the periodogram, there is significant non-random structure in the
noise continuum, particularly for trial periods longer than 1 day. Our
interpretation of this phenomenon is that it is a by-product of a few
unique features of the transit signals we are dealing with. Firstly
the duration of the transits, and hence the width of the boxes fitted
by the BLS algorithm, is much shorter with respect to both the orbital
period and the WASP survey cadence than for transits of main sequence
stars. As a consequence the trial bins will contain many fewer data
points than for a main sequence transit search, particularly at longer
trial periods. Secondly as the transit signals are so deep compared to
the main sequence case, they are less prone to being ``washed out''
when the light curve is folded on an incorrect trial period and the
in-transit points spread across all orbital phases. In the case
illustrated the Earth-sized body will cause transits with a depth of
$\sim 70$\%. Even in a trial bin with, for example, 1 in-transit and
10 out-of-transit data points the presence of the single in-transit
point would drag down the mean light level in the bin by more than
6\%, which could be sufficient to be regarded as a significant
detection.

\begin{figure} 
\centering
  \psfig{file=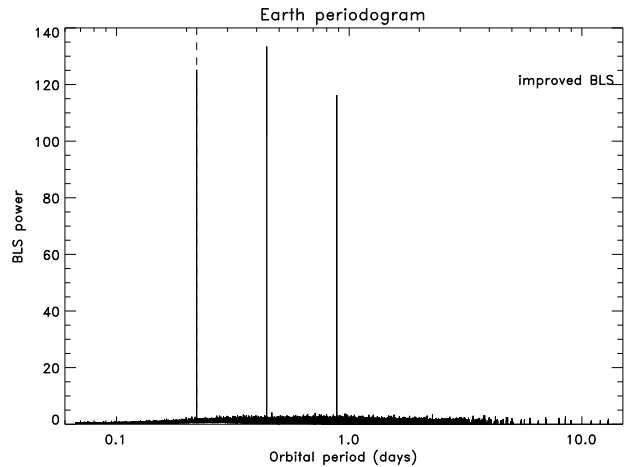,width=0.5\textwidth} 
\caption{BLS power spectrum as in Figure \ref{old_prdg} obtained with the
    improved BLS routine. For the same transit signal we achieve
    higher {\em S/N} values and higher statistical significance for the
    detection. The strong aliasing and the noise structure seen in
    Figure \ref{old_prdg} is much reduced.}
\label{new_prdg}
\end{figure}
\noindent 

To address this issue we modified equation~\ref{sn_unmod} as follows

\begin{equation}
\mathcal{S}_j'=\frac{\delta_j}{\sqrt{{\rm Var}(\delta_j)+\hat{d}^2_j}}
\end{equation}

where $\hat{d}^2_j$ is the mean-square deviation of the data points
within the bin about their mean. This modification will have the
effect of strongly reducing the computed signal-to-noise of bins which
contain a mix of in-transit and out-of-transit data points. The
magnitude of this down-weighting will tend to increase as the depth of
the transit signal increases, but only in the cases in which the trial
period does not match the true period (or an integer multiple or
fraction). Where a bin contains just in-transit points the
down-weighting will generally be small for all transit depths.
Figure~\ref{new_prdg} shows the BLS periodogram computed using this
modified prescription for the signal-to-noise, for the same transit as
in Figure \ref{old_prdg}.

\subsection{False detection rate}
Automated searches for weak signals in noisy data are inherently
susceptible to ``false alarms'', whereby a chance alignment of noise
fluctuations in the data are misinterpreted by the search algorithm as
a evidence likely detection of the signal being hunted. It is useful
therefore to be able to define a filter which can be applied
automatically to weed out these false detections. We achieved this by
constructing synthetic light curves based on the time-sampling of
sample WASP light curves which contain pure white-noise, but no
simulated transit signal. We then computed BLS periodograms for these
light curves in the same manner as for those containing simulated
transits.

\begin{figure}
\centering
\psfig{file=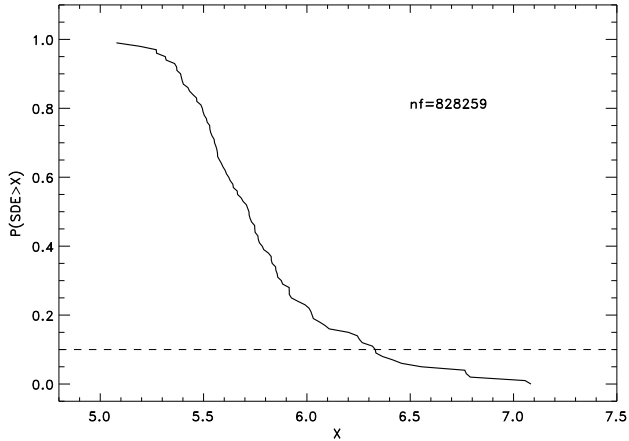,width=0.5\textwidth}
\caption{Probability Distribution Function (PDF) for the
  modified-optimised BLS routine. The dashed-line shows the detection
  threshold 6.3{\em SDE} for a 10\% noise contribution.}
\label{PDF_new}
\end{figure}

\citet{Kovacs02} define a useful metric for assessing the likely
significance of a peak in a BLS periodogram, which they refer to as
the Signal Detection Efficiency (SDE):

\[
\mathrm {SDE}=\frac{\mathcal {S}_{\rm peak} -
  \bar{\mathcal{S}}}{\sigma_{\mathcal{S}}}
\]

where $\mathcal{S}_{\rm peak}$ is the height of the peak, and
$\bar{\mathcal{S}}$ and $\sigma_{\mathcal{S}}$ are measures of the
mean level and scatter in the noise continuum of the periodogram. We
computed the SDE of the highest peak in each of the synthetic,
transitless light curves. The cumulative distribution function of
these SDE measures over the whole sample is plotted in
Figure~\ref{PDF_new}. This distribution function allowed us to define
a threshold SDE below which we could automatically discount a
detection as likely to be a false alarm. As the size of the sample of
WDs observed by WASP is fairly small, just a few hundred, we
chose a relatively generous threshold (SDE$_{\rm thresh}=6.3$) which
would allow through around 10\% of false detections. For larger
surveys a more strict threshold might be necessary to avoid being
swamped by false detections.

\noindent

\begin{figure}
\centering
\psfig{file=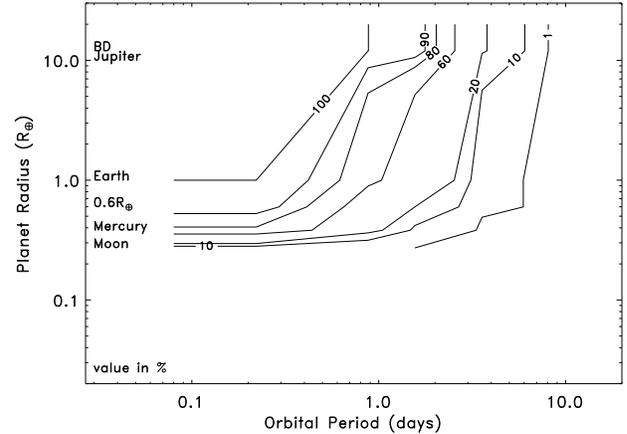,width=0.5\textwidth}
\psfig{file=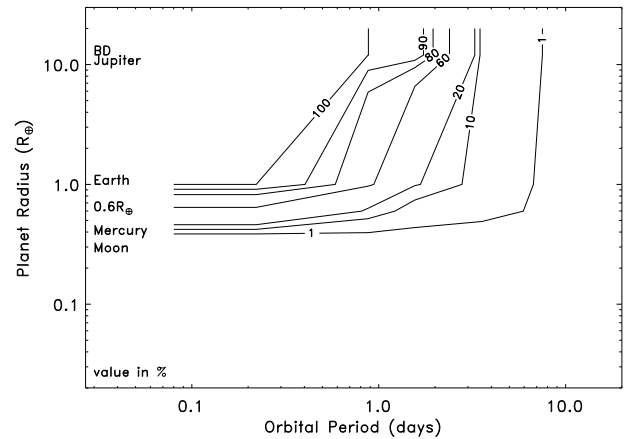,width=0.5\textwidth}
\psfig{file=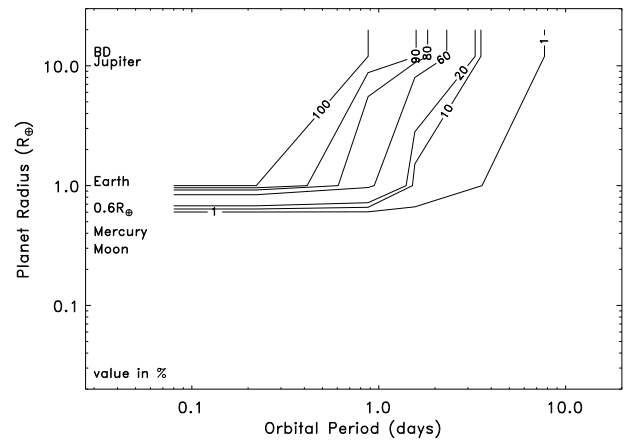,width=0.5\textwidth}
\caption{Recovery rate of simulated transits of a WD of magnitude
  $\textrm{V}\simeq12$ (top), $\textrm{V}\simeq13$ (middle) and
  $\textrm{V}\simeq15$ (bottom). The frequency contours are expressed in
  percentage values.}
\label{det_rate}
\end{figure}

\begin{table*}
 \smallskip
 \begin{center}
   \caption{Recovery rate of simulated transits of a bright WD
     ($\textrm{V}\simeq12$). Results are shown for synthetic light curves containing white and
     red noise. In both cases $f_{\rm det}$ is the 
     fraction of cases in which the highest peak in the
     periodogram satisfies our period matching criteria and has an ${\rm
     SDE}>6.3$, and $f_{\rm bt}$ is the fraction which match the period criteria
     but have ${\rm SDE}<6.3$. $f_{\rm filt}$ is the fraction of cases in which the synthetised light curves pass the
     requirements for a minimum number of transits ($>5$) and data points ($>5$) in
     transit. Dashes indicate cases in which the companion would be
     tidally disrupted within the WD's Roche radius.  
   }

\begin{tabular}{ccccrrrrrcrr}
\hline
\noalign{\smallskip}
&&&&&&&\multicolumn{2}{c}{White noise}&&\multicolumn{2}{c}{red noise}\\
\noalign{\smallskip}
\cline{8-12}
\noalign{\smallskip}
Size&&$R_{\rm pl}$ & $\delta_{\rm tr}$ &$P$ & $D_{\rm tr}$ & $f_{\rm filt}$ & $f_{\rm det}$ &
$f_{\rm bt}$ &  & $ f_{\rm det}$ &$f_{\rm bt}$ \\
&&\small{\Re} & (\%) &\small{(days)}&\small{(min)}&(\%)&(\%)&(\%)&&(\%)&(\%) \\
\noalign{\smallskip}
\hline
\hline
\noalign{\smallskip}
BD-Gas Giant       & & 10.0    & 100    &   0.08     &   5.65 & 100     & 100 &      0   & & 100 &     0 \\
                             & &            &          &   0.22    &   7.93 & 100     & 100 &      0   & & 100 &     0  \\
                         & &            &          &   0.87     &  12.55 & 100     & 100 &      0   & & 100 &     0  \\
                         & &            &          &   1.56     &  15.22 & 100     &   99 &      1   & &   98  &     2  \\
                         & &            &             &   3.57     &  20.05 &  69     &   24 &   25   & &    21  &  32 \\
                         & &            &             &   8.30     &  26.56 & 0.1     &    0  &      0   & &      0  &    0  \\
                         & &            &             &  14.72 &  32.15 &   0     &    0  &      0   & &      0  &    0  \\
\\
Earth     & & 1.0    &  49    &   0.08    &   $-$      & $-$ & $-$ &  $-$ & & $-$ & $-$ \\
             & &            &             &   0.22     &   1.70     & 100    & 100    &    0    & & 100 &     0  \\
             & &            &             &   0.87     &   2.65     &  96     &  76     &    5        & &   67 &   10  \\
             & &            &             &   1.56     &   3.21     &  71     &  48    &  19        & &   37 &   30  \\
             & &            &             &   3.57     &   4.23     &  14     &   5     &    7        & &    2 &    4  \\
             & &            &             &   8.30     &   5.61     &   0     &   0     &    0       & &    0 &    0  \\
             & &            &             &  14.72 &   6.82     &   0     &   0     &    0        & &    0 &    0  \\
\\
0.6\Re  & & 0.6    &  18    &   0.08     &   $-$  & $-$ & $-$ & $-$ & & $-$ & $-$  \\
             & &            &             &   0.22     &   1.42 & 100    &  98    &    2      & & 96 &     4  \\
             & &            &             &   0.87     &   2.21 &  87     &  44    &  14      & & 41 &     6  \\
             & &            &             &   1.56     &   2.68 &  51     &  31     &  17     & & 20 &   19  \\
             & &            &             &   3.57     &   3.53 &   7         &   3      &    2     & &   2 &     1  \\
             & &            &             &   8.30     &   4.67 &   0         &   0      &    0     & &   0 &     0  \\
             & &            &             &  14.72 &   5.66 &   0         &   0      &    0      & &   0 &     0  \\
\\
Mercury    & & 0.34    &  5.7    &    0.08 &   $-$  & $-$ & $-$ & $-$ & & $-$ & $-$  \\
                 & &            &             &    0.22 &   1.25 & 100    &  86     &    6    & &  78 &    2  \\
                 & &            &             &    0.87 &   1.98 &  78     &  25     &  26      & &  24 &  19  \\
                 & &            &             &    1.56 &   2.40 &  40     &  12     &  14        & &    8 &    9  \\
                 & &            &             &    3.57 &   3.17 &   4         &    2     &    0        & &    0 &    0   \\
                 & &            &             &    8.30 &   4.20 &   0         &    0     &    0     & &    0 &    0  \\
                 & &            &             &  14.72 &   5.10 &   0         &    0     &    0       & &    0 &    0  \\
\\
Moon    & & 0.27    &  3.6    &    0.08 &    $-$ & $-$ & $-$ & $-$  & & $-$ & $-$  \\
             & &            &             &    0.22 &   1.19 & 100    &  38     &  13       & &    4 &  26  \\
             & &            &             &    0.87 &   1.87 &  74     &  12     &  24       & &    1 &  30  \\
             & &            &             &    1.56 &   2.27 &  35     &    4     &  33       & &    1 &  37  \\
             & &            &             &    3.57 &   2.99 &   3         &    0     &    0      & &    0 &    0  \\
             & &            &             &    8.30 &   3.96 &   0         &    0     &    0      & &    0 &    0  \\
             & &            &             &  14.72 &   4.79 &   0         &    0     &    0          & &    0 &    0  \\
\noalign{\smallskip}
\hline
\hline
\label{tab12}
\end{tabular}

\end{center}
\end{table*}

\subsection{Recovery rates of synthetic transits}

Figure~\ref{det_rate} and Tables~\ref{tab12}, \ref{tab13} and
\ref{tab15} summarise our recovery rate for simulated transit signals
injected into synthetic light curves of WDs of magnitudes
$\textrm{V}\simeq12$, $\textrm{V}\simeq13$ and $\textrm{V}\simeq15$
respectively. We regard as a match any trial in which the most
significant detected period is within 1\% of being an integer fraction
or multiple from $1/5\times$ to $5\times$ the injected transit signal.

We have attempted to separate out the various factors which can affect
the efficiency of detection of these transit signals. When generating
each synthetic light curve we can readily assess {\it a priori}
whether it will fail the tests requiring a minimum number of
individual transits and in-transit data points. We list in
Tables~\ref{tab12}, \ref{tab13} and \ref{tab15} the fraction $f_{\rm
  filt}$ which pass these two tests. It is evident from these tables
that these requirements alone render transiting companions essentially
undetectable at our longest trial periods (8.30 and 14.72 days) in a
WASP-like survey; the transits are too short in duration and too
infrequent to be adequately sampled. For companions around 1\Re\ and
larger however there is a good chance of detection out to periods of
around 4 days, at least in principle.

The table also shows the impact of adding representative photometric
noise on the detection rates ($f_{\rm det}$). For the idealised
photon-noise-limited case objects as small as Mercury could be
detected to periods of around 1.5\,d, and the Moon for periods less
than 1\,d. Once the impact correlated instrumental noise (red noise)
is added, Moon-sized companions become almost undetectable, though the
recovery rates for larger bodies, particularly in short-period orbits,
remains encouraging.

Our key conclusion from these simulations is that for the case of
transits of WDs, the degree of photometric precision delivered
by a survey is of somewhat secondary importance compared to a high
cadence and continuous coverage. For planet-sized bodies individual
transits will be quite deep and readily detectable in data of moderate
photometric quality, however it is the short duration of the transits
that is the main factor limiting the transit detection rate in surveys
optimised for main sequence stars.

\section{Searching for transit signals in WASP survey data}

Encouraged by the results of our simulations we selected a sample of
WDs, which have been routinely monitored by WASP through the
2004--2008 observing seasons, and performed a systematic search for
eclipsing and transiting substellar and planetary companions. We
selected the sample by cross-correlating the catalogue of WASP objects
for which more than 600 data points are available with the McCook $\&$
Sion catalogue \citep{WDcat}. The resulting sample of 194 WDs with
magnitude $\textrm{V}<15$ is presented in Table~4.

We searched the sample for transits and eclipses using our
implementation of the BLS algorithm, searching periods ranging from 2
hours to 15 days. In addition we have also inspected each of the
individual light curves by eye. In both searches we found no evidence
for any transiting and eclipsing companions within the period range
searched in this study. We have used this null result together with
the results of our simulations to estimate an upper-limit to the
frequency of such close companions for the sample of WDs
considered in this study.

\subsection{Limits on frequency of companions to WDs}

In order to estimate an upper limit to the frequency of close
substellar and planetary companions to WDs, we used the
detection limits derived from our simulations and the results obtained
from the analysis of the sample of 194 WDs. We first used a
binomial distribution to describe the probability $\mathcal{P}(n;N,f)$
of finding $n$ transiting companions for a given sample of $N$ stars,
with a true companion frequency $f$ (e.g.. see \citealt{McCarthy04};
and Appendix of \citealt{Burgasser03}) as follows:

\begin{equation}
\mathcal{P}(n;N,f) = \frac{N!}{n!(N-n)!}f^{n}(1-f)^{N-n}
\label{distribution}
\end{equation}
\noindent
When the two quantities $N$ and $n$ are known equation
\ref{distribution} can be used to derive the distribution
($\mathcal{P}_{1}$) describing the probability of $f$, where $f$ is the
frequency of transiting companions. The probability
$\mathcal{P}_{1}(f;n,N)$ is proportional to $\mathcal{P}(n;N,f)$ for $f$ in the
interval $[0,1]$. We obtain $\mathcal{P}_{1}$ by normalising :
\begin{equation}
\int\limits_{0}^{1}\mathcal{P}_{1}(f;n,N)\ df\ = 1
\label{p1}
\end{equation}
\noindent
which yields $\mathcal{P}_{1}=(N+1)\mathcal{P}$.

\begin{figure} 
\centering
  \psfig{file=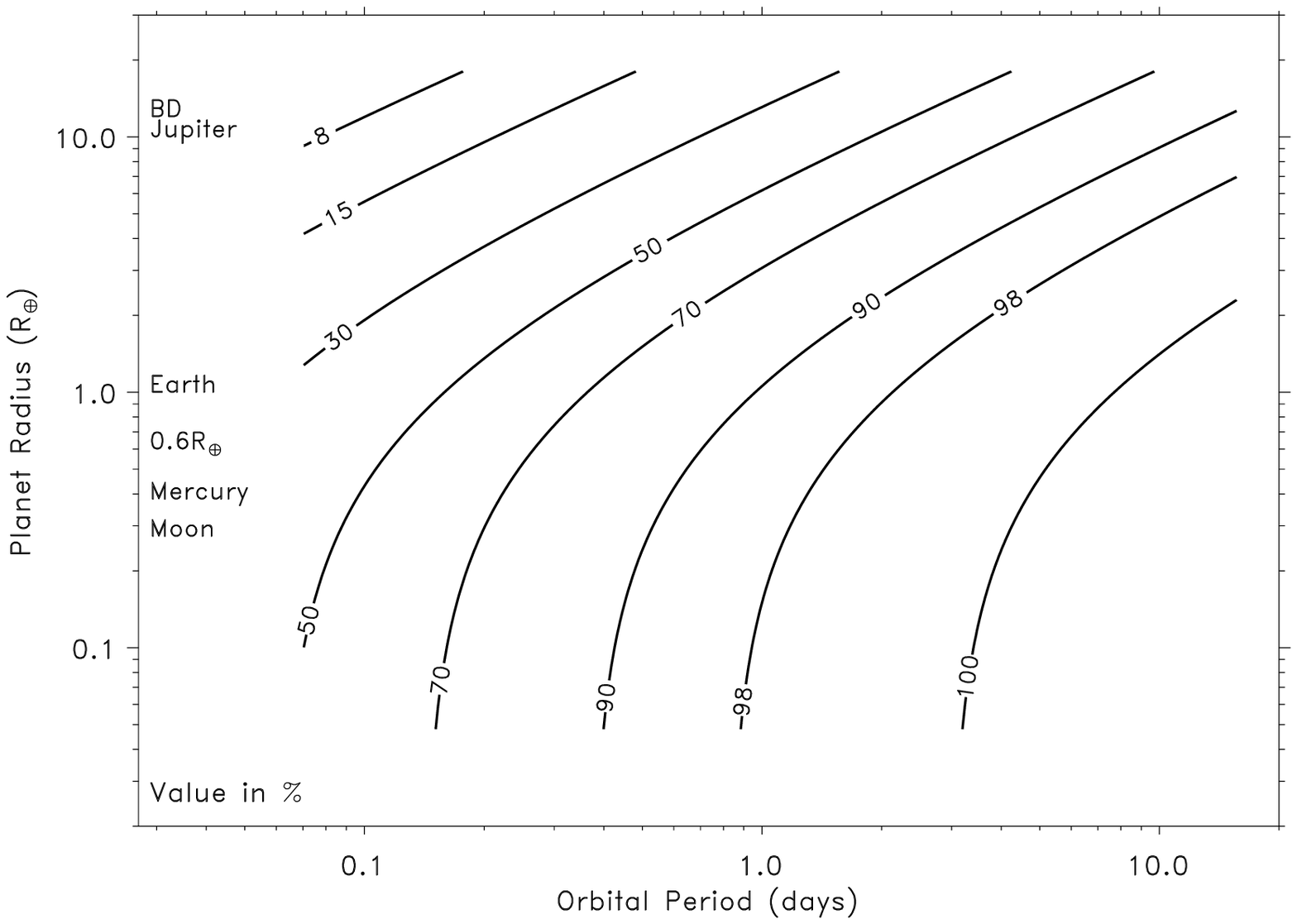,width=0.5\textwidth}\\
  \psfig{file=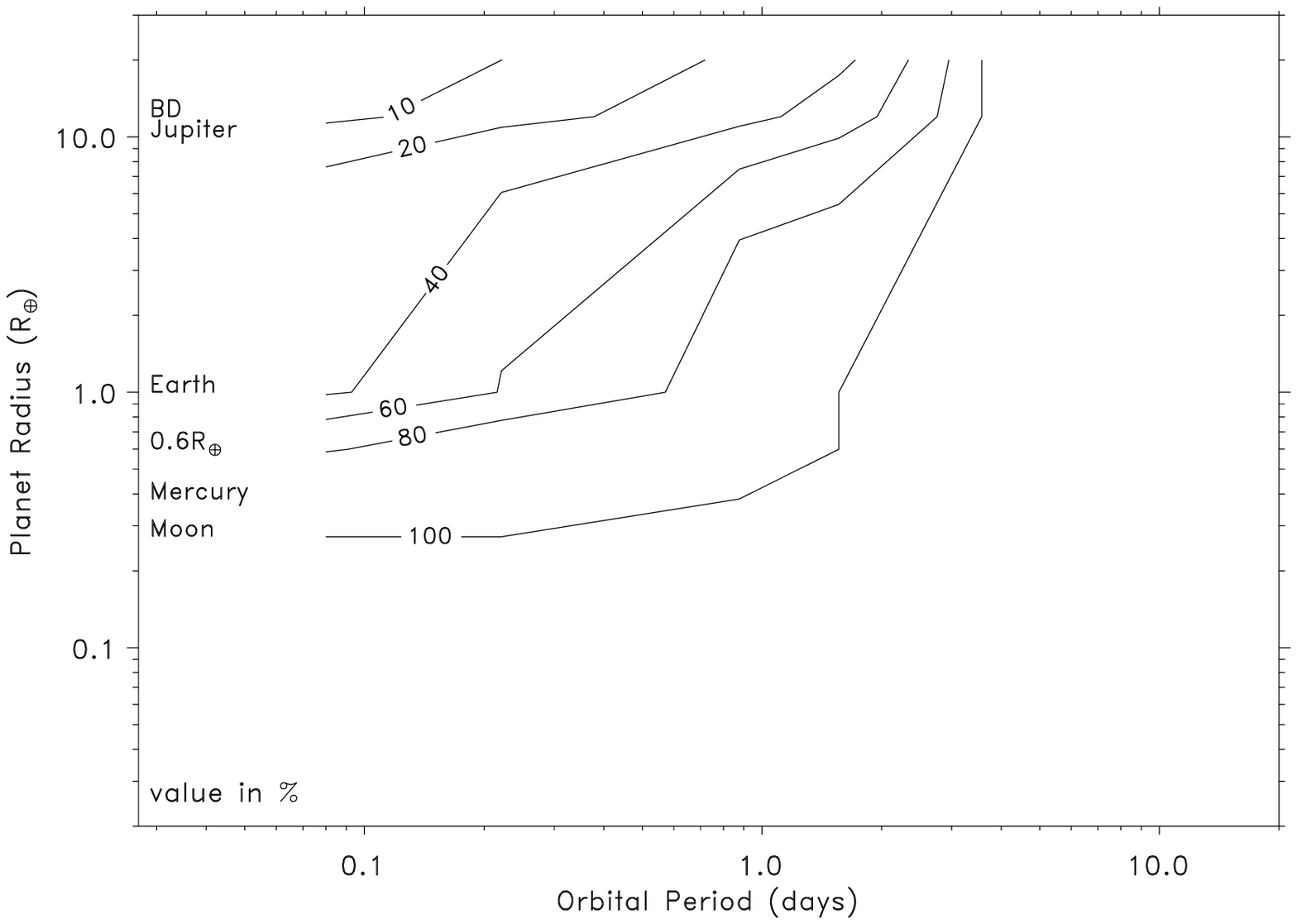,width=0.5\textwidth}
  \caption{Top panel: upper-limit on companion frequency inferred from
    a null detection in a survey sample of size $N=194$ assuming
    perfect detectability of transits across the parameter space.
    Lower panel: limits in the same sample folding in the
    detectability of transiting systems in a WASP-like survey. In both
    panels the frequency contours are expressed in percentage values.}
\label{limits}
\end{figure}

Although our complete sample numbers $N=194$ stars, we have already
established that even if all of these have companions only a fraction
$p_{\rm tr}(R_p, P)$ will exhibit a transit, and of those which do
exhibit a transit only a fraction $p_{\rm det}(R_p, P)$ would be
detectable in a WASP-like survey. Both of these factors will act to
reduce the total number of transiting companions detected in the
survey, or in the case of a null result will tend to weaken the
constraints that can be placed on true companion frequency by such a
survey. To incorporate these factors we modified our effective sample
size as:

\[
N'=N\times p_{\rm tr}(R_p, P)\times p_{\rm det}(R_p, P)
\]

and used this in Eq.~\ref{p1}, which we integrated to find the
limiting companion frequency $f_{\rm lim}$ that encloses 95\% of the
probability distribution. Figure~\ref{limits} (top panel) shows the
upper-limit on companion frequency for a null detection in a
``perfect'' survey in which $p_{\rm det}=1$ and a sample size $N=194$.
In such a survey the detectability of companions is limited solely by
the intrinsic probability of them transiting their host.

\noindent

To factor in the efficiency of detection of transits in a WASP-like
survey we need to determine a representative $p_{\rm det}(R_p, P)$.
Our simulations were performed at only three specific host-star
magnitudes, whereas the distribution of the magnitudes of the stars in
our sample is of course a continuum (covering the range $\textrm{V}\sim9-15$).
We therefore combine the three magnitude-specific $p_{\rm det}$ maps
(Figure~\ref{det_rate}) into a single map by
interpolating/extrapolating according to the magnitude of each object
in our sample and combining these to form an averaged map which can be
folded in to our calculation of the upper-limits. The resulting limits
corresponding to the 95\% of the integrated probability, are shown in
in the lower panel of Figure~\ref{limits}. Our results show that for
rocky bodies smaller than the size of Mercury no useful upper limits
to the frequency of companions to WDs can be found, and that
for Earth-sized companions only weak constraints can be imposed.
However, it does suggest that objects the size of brown dwarfs or gas
giants with orbital periods P$<0.1-0.2$~days must be relatively rare
(upper limit of $\sim$10\%).

\section{Conclusion}

We have used a modified version of the Box Least Squares algorithm to
investigate the detection limits for substellar and planetary
companions to WDs achievable using data already available in
the WASP photometric survey. Our simulations proved extremely
encouraging, suggesting that planetary bodies as small as Mercury at
small orbital radii can be detected with good photometric data even in
the presence of red noise. For smaller bodies red noise in the
light curves becomes increasingly problematic, while for bodies with
larger orbital periods, the absence of significant numbers of
in-transit points, significantly decreases the detection sensitivity.
 
Application of our modified BLS algorithm to search for companions to
WDs in a sample of 194 stars in the magnitude range
$\textrm{V}\sim9$--15, available in the WASP archive, did not reveal any
eclipsing or transiting substellar or planetary companions. Visual
inspection of individual light curves for the WDs in our
sample confirmed the absence of significant periodic dropouts in the
WASP data. We have used the non-detection of planetary companions to
the WDs in our sample together with the estimated detection
sensitivities determined from our simulations, to place upper limits
to the frequency of substellar and planetary companions to white
dwarfs. While no useful limits can be placed on the likely frequency
of Mercury-sized or smaller companions, and only weak constraints on
the frequency of Earth-sized objects in the closest orbits, slightly
stronger constraints can be placed on the frequency of larger bodies
in very short-period orbits. For example, brown dwarfs and gas giants
radius $\approx \Rjup$ with periods $<0.1-0.2$~days, similar to the
known WD$+$BD binary WD$0137-349$ \citep{Maxted06}, must certainly be
relatively rare ($\la10\%$). Of course, this limit needs to be
compared with those derived from other sources, e.g. infra-red sky
surveys. For example, \citet{Farihi05} estimated that $<0.5\%$ of
WDs have L dwarf companions, while Steele et al. (in prep.)
tentatively suggests the fraction of unresolved brown dwarf companions
(including T dwarfs) may be slightly higher, between $1-2\%$. From
Spitzer photometry \citet{Farihi08a} suggests that $<4\%$ of white
dwarfs have unresolved substellar companions $>10\Mjup$, although the
limits at lower masses (e.g. $<6\Mjup$) are considerably weak.

To place more stringent constraints on close substellar and gas giant
companions to WDs, and similarly stringent constraints on
Earth size bodies in close orbital separations likely requires
significantly larger WD samples. In addition, our simulations
and analysis of WD light curves in the WASP archive suggests
the degree of photometric precision is of somewhat secondary
importance compared to a high cadence and continuous coverage. The
short duration of eclipses and transits of WDs
($\approx5-20$~mins for companions radius~$\approx \Rjup$;
$\approx1-5$~mins for terrestrial bodies) compared to the $\approx
8$~min cadence of WASP observations, appears to be the main factor
limiting the transit detection rate in a survey optimised for
planetary transits of main sequence stars.

Future surveys such as Pan-STARRS and LSST will be capable of
detecting tens of thousands of WDs. However, we emphasise
that observations of high cadence and long baseline are of greatest
benefit when attempting to detect the signature of close, eclipsing
and transiting substellar and planetary companions to WDs.
Space missions such as {\it COROT}, {\it Kepler} (see
\citealt{DiStefano}) and, especially, {\it PLATO} may therefore be
better suited to a survey of WDs as they deliver
uninterrupted coverage at high cadence and exquisite photometric
precision ($\sim 10^{-4}-10^{-5}$) and could at least in principle
detect the transits of asteroid-sized bodies across a WD.

\section{Acknowledgments}
FF acknowledges funding from the European Commission under the Marie
Curie Host Fellowship for Early Stage Research Training SPARTAN,
Contract No MEST-CT-2004-007512, University of Leicester, UK. MRB acknowledges the
support of an STFC Advanced Fellowship during part of this research.
The WASP Consortium consists of astronomers primarily from the Queen's
University Belfast, Keele, Leicester, The Open University, and St
Andrews, the Isaac Newton Group (La Palma), the Instituto de
Astrofisica de Canarias (Tenerife) and the South African Astronomical
Observatory. The SuperWASP-N and WASP-S Cameras were constructed and
operated with funds made available from Consortium Universities and
the UK's Science and Technology Facilities Council. WASP-South is
hosted by the South African Astronomical Observatory (SAAO) and we are
grateful for their support and assistance. The author also thank
Professor Andrew Collier Cameron and the anonymous referee for helpful comments on this paper.

\newpage

\begin{table*}
  \caption{Recovery rate of simulated transits of a $\textrm{V}\simeq13$ WD}
  \label{tab13}
\begin{tabular}{ccccrrrrrcrr}
\hline
\noalign{\smallskip}
&&&&&&&\multicolumn{2}{c}{White noise}&&\multicolumn{2}{c}{red noise}\\
\noalign{\smallskip}
\cline{8-12}
\noalign{\smallskip}
Size&&$R_{\rm pl}$ & $\delta_{\rm tr}$ &$P$ & $D_{\rm tr}$ & $f_{\rm filt}$ & $f_{\rm det}$ &
$f_{\rm bt}$ &  & $ f_{\rm det}$ &$f_{\rm bt}$ \\
&&\small{\Re} & (\%) &\small{(days)}&\small{(min)}&(\%)&(\%)&(\%)&&(\%)&(\%) \\
\noalign{\smallskip}
\hline
\hline
\noalign{\smallskip}
BD-Gas Giant   	& & 10.0 & 100 &   0.08  &   5.65 & 100    & 100  &    0  & & 100 &   0  \\
             	& &      &     &   0.22  &   7.93 & 100    & 100  &    0  & & 100 &   0  \\
             	& &      &     &   0.87  &  12.55 & 100    & 100  &    0  & & 100 &   0  \\
             	& &      &     &   1.56  &  15.22 & 100    &  95  &    2  & &  98 &   0  \\
             	& &      &     &   3.57  &  20.05 &  69    &  24  &   25  & &   6 &  31  \\
             	& &      &     &   8.30  &  26.56 & 0.1    &   0  &    5  & &   0 &   7  \\
             	& &      &     &  14.72  &  32.15 &   0    &   0  &    0  & &   0 &   1  \\
\\							  
Earth		& & 1.0 &  49  &   0.08  &   $-$  & $-$    & $-$  &  $-$  & & $-$ & $-$ \\
          	& &     &      &   0.22  &   1.70 & 100	   & 100  &    0  & & 100 &    0  \\
          	& &     &      &   0.87  &   2.65 &  96    &  74  &   11  & &  64 &    6  \\
          	& &     &      &   1.56  &   3.21 &  71    &  37  &   30  & &  21 &   17  \\
          	& &     &      &   3.57  &   4.23 &  14    &   2  &   34  & &   3 &   36  \\
          	& &     &      &   8.30  &   5.61 &   0    &   0  &    4  & &   0 &    5  \\
          	& &     &      &  14.72  &   6.82 &   0    &   0  &    0  & &   0 &    2  \\
\\							  
0.6\Re		& & 0.6	&  18  &   0.08  &   $-$  & $-$    & $-$  & $-$   & & $-$ & $-$  \\
        	& &     &      &   0.22  &   1.42 & 100	   &  84  &    1  & &  55 &   29  \\
        	& &     &      &   0.87  &   2.21 &  87    &  42  &   11  & &  16 &   26  \\
        	& &     &      &   1.56  &   2.68 &  51    &  15  &   20  & &   4 &   53  \\
        	& &     &      &   3.57  &   3.53 &   7    &   3  &   35  & &   2 &   37  \\
        	& &     &      &   8.30  &   4.67 &   0    &   0  &    5  & &   0 &    4  \\
        	& &     &      &  14.72	 &   5.66 &   0    &   0  &    2  & &   0 &    0  \\
\\			       	 	 		  
Mercury		& & 0.45 &  10 &    0.08 &   $-$  & $-$    & $-$  & $-$  & & $-$ & $-$  \\
        	& &      &     &    0.22 &   1.25 & 100    &  22  &  19  & &   0 &    8  \\
        	& &      &     &    0.87 &   1.98 &  78    &  16  &  15  & &   0 &   26  \\
        	& &      &     &    1.56 &   2.40 &  40    &  13  &  22  & &   0 &   51  \\
        	& &      &     &    3.57 &   3.17 &   4    &   1  &  35  & &   0 &   35  \\
        	& &      &     &    8.30 &   4.20 &   0    &   0  &   4  & &   0 &    4  \\
        	& &      &     &   14.72 &   5.10 &   0    &   0  &   2  & &   0 &    0  \\
\\							  
Moon		& & 0.27 &  3.6	&    0.08 &  $-$ & $-$     & $-$  & $-$  & & $-$ & $-$  \\
        	& &      &     	&    0.22 &  1.19 & 100	   &  12  &   8  & &   0 &    8  \\
        	& &      &     	&    0.87 &  1.87 &  74    &   8  &  15  & &   0 &   26  \\
        	& &      &     	&    1.56 &  2.27 &  35    &   3  &  23  & &   0 &   50  \\
        	& &      &     	&    3.57 &  2.99 &   3    &   0  &  35  & &   0 &   35  \\
        	& &      &     	&    8.30 &  3.96 &   0    &   0  &   4  & &   0 &    4  \\
        	& &      &     	&  14.72  &  4.79 &   0    &   0  &   2  & &   0 &    0  \\
\noalign{\smallskip}								   
\hline										   
\hline
\end{tabular}
\end{table*}

\begin{table*}
  \caption{Recovery rate of simulated transits of a $\textrm{V}\simeq15$ WD}
  \label{tab15}
\begin{tabular}{ccccrrrrrcrr}
\hline
\noalign{\smallskip}
&&&&&&&\multicolumn{2}{c}{White noise}&&\multicolumn{2}{c}{red noise}\\
\noalign{\smallskip}
\cline{8-12}
\noalign{\smallskip}
Size&&$R_{\rm pl}$ & $\delta_{\rm tr}$ &$P$ & $D_{\rm tr}$ & $f_{\rm filt}$ & $f_{\rm det}$ &
$f_{\rm bt}$ &  & $ f_{\rm det}$ &$f_{\rm bt}$ \\
&&\small{\Re} & (\%) &\small{(days)}&\small{(min)}&(\%)&(\%)&(\%)&&(\%)&(\%) \\
\noalign{\smallskip}
\hline
\hline
\noalign{\smallskip}
BD-Gas Giant   	& & 10.0 & 100 &   0.08  &   5.65 & 100    & 100  &    0  & & 100 &   0  \\
             	& &      &     &   0.22  &   7.93 & 100    & 100  &    0  & & 100 &   0  \\
             	& &      &     &   0.87  &  12.55 & 100    & 100  &    0  & & 100 &   0  \\
             	& &      &     &   1.56  &  15.22 & 100    &  98  &    0  & &  91 &   3  \\
             	& &      &     &   3.57  &  20.05 &  69    &   4  &   33  & &   8 &  44  \\
             	& &      &     &   8.30  &  26.56 & 0.1    &   0  &    5  & &   0 &   8  \\
             	& &      &     &  14.72  &  32.15 &   0    &   0  &    0  & &   0 &   2  \\
\\							  
Earth		& & 1.0 &  49  &   0.08  &   $-$  & $-$    & $-$  &  $-$  & & $-$ & $-$ \\
          	& &     &      &   0.22  &   1.70 & 100	   & 100  &    0  & & 100 &    0  \\
          	& &     &      &   0.87  &   2.65 &  96    &  65  &   11  & &  66 &    8  \\
          	& &     &      &   1.56  &   3.21 &  71    &  37  &   30  & &   6 &   43  \\
          	& &     &      &   3.57  &   4.23 &  14    &   2  &   34  & &   1 &   49  \\
          	& &     &      &   8.30  &   5.61 &   0    &   0  &    4  & &   0 &    7  \\
          	& &     &      &  14.72  &   6.82 &   0    &   0  &    0  & &   0 &    1  \\
\\							  
0.6\Re		& & 0.6	&  18  &   0.08  &   $-$  & $-$    & $-$  & $-$   & & $-$ & $-$  \\
        	& &     &      &   0.22  &   1.42 & 100	   &  10  &   18  & &   0 &    9  \\
        	& &     &      &   0.87  &   2.21 &  87    &   7  &   17  & &   0 &   27  \\
        	& &     &      &   1.56  &   2.68 &  51    &   4  &   44  & &   0 &   52  \\
        	& &     &      &   3.57  &   3.53 &   7    &   1  &   50  & &   0 &   36  \\
        	& &     &      &   8.30  &   4.67 &   0    &   0  &    7  & &   0 &    4  \\
        	& &     &      &  14.72	 &   5.66 &   0    &   0  &    1  & &   0 &    0  \\
\\			       	 	 		  
Mercury		& & 0.45 &  10 &    0.08 &   $-$  & $-$    & $-$  & $-$  & & $-$ & $-$  \\
        	& &      &     &    0.22 &   1.25 & 100    &   5  &   8  & &   0 &    8  \\
        	& &      &     &    0.87 &   1.98 &  78    &   5  &  18  & &   0 &   26  \\
        	& &      &     &    1.56 &   2.40 &  40    &   3  &  45  & &   0 &   50  \\
        	& &      &     &    3.57 &   3.17 &   4    &   0  &  49  & &   0 &   35  \\
        	& &      &     &    8.30 &   4.20 &   0    &   0  &   6  & &   0 &    4  \\
        	& &      &     &   14.72 &   5.10 &   0    &   0  &   1  & &   0 &    0  \\
\\							  
Moon		& & 0.27 &  3.6	&    0.08 &  $-$ & $-$     & $-$  & $-$  & & $-$ & $-$  \\
        	& &      &     	&    0.22 &  1.19 & 100	   &   4  &   7  & &   0 &    8  \\
        	& &      &     	&    0.87 &  1.87 &  74    &   4  &  18  & &   0 &   26  \\
        	& &      &     	&    1.56 &  2.27 &  35    &   1  &  43  & &   0 &   49  \\
        	& &      &     	&    3.57 &  2.99 &   3    &   0  &  49  & &   0 &   35  \\
        	& &      &     	&    8.30 &  3.96 &   0    &   0  &   6  & &   0 &    4  \\
        	& &      &     	&  14.72  &  4.79 &   0    &   0  &   0  & &   0 &    0  \\
\noalign{\smallskip}								   
\hline										   
\hline
\end{tabular}
\end{table*}

\newpage
\onecolumn

\begin{center}
\tablefirsthead{
\hline
\multicolumn{1}{c}{1SWASP} && \multicolumn{1}{c}{WD} &&&
\multicolumn{1}{c}{V} && \multicolumn{1}{c}{N}\\
\multicolumn{1}{c}{} && \multicolumn{1}{c}{} &&&
\multicolumn{1}{c}{(WASP)} && \multicolumn{1}{c}{}\\
}
\tablehead{
\multicolumn{4}{|l|}{\small\sl continued from previous page}\\
\hline
\multicolumn{1}{c}{1SWASP} && \multicolumn{1}{c}{WD} &&&
\multicolumn{1}{c}{V} && \multicolumn{1}{c}{N}\\
\multicolumn{1}{c}{} && \multicolumn{1}{c}{} &&&
\multicolumn{1}{c}{(WASP)} && \multicolumn{1}{c}{}\\
\hline}
\tabletail{
\hline
\multicolumn{4}{|r|}{\small\sl continued on next page}\\
\hline}
\tablelasttail{\hline}
\begin{supertabular}{cccccccc}\hline
J000007.24+295700.6   & & 2357+296    &  &&12.24 &&15825  \\
J000331.62$-$164358.4 & & 0000$-$170  &  &&14.87 &&10446  \\
J000732.24+331727.7   & & 0004+330    &  &&13.95 &&21067  \\
J000818.17+512316.7   & & 0005+511    &  &&13.42 && 4876  \\
J002130.72$-$262611.0 & & 0018$-$267  &  &&13.92 && 7305  \\
J003112.96$-$271253.7 & & 0028$-$274  &  &&14.99 && 6961  \\
J003145.95+571817.1   & & 0029+571    &  &&10.50 && 2511  \\
J003353.90$-$270823.6 & & 0031$-$274  &  &&14.30 && 7293  \\
J003952.15+313229.3   & & 0037+312    &  &&15.03 && 9429  \\
J004121.46+555009.1   & & 0038+555    &  &&13.47 && 6202  \\
J005317.46$-$325956.6 & & 0050$-$332  &  &&13.45 && 9316  \\
J005340.53+360118.4   & & 0050+357    &  &&14.54 &&14264  \\
J011011.78+270104.8   & & 0107+267    &  &&15.61 && 4702  \\
J011018.59$-$340025.5 & & 0107$-$342  &  &&14.27 &&10151  \\
J011211.65$-$261327.7 & & 0109$-$264  &  &&13.14 &&10038  \\
J011547.45$-$240651.0 & & 0113$-$243  &  &&15.03 &&29846  \\
J012942.57+422817.1   & & 0126+422    &  &&13.51 &&18136  \\
J013856.85+152742.5   & & 0136+152    &  &&14.37 && 2251  \\
J014754.80+233943.8   & & 0145+234    &  &&14.19 && 4653  \\
J015202.95+470005.5   & & 0148+467    &  &&12.08 && 4362  \\
J020253.98$-$165303.5 & & 0200$-$171  &  &&11.41 && 8756  \\
J021255.35+170356.5   & & 0210+168    &  &&14.32 && 3082  \\
J021616.34+395125.5   & & 0213+396    &  &&14.11 && 7608  \\
J021733.49+570647.3   & & 0214+568    &  &&13.29 && 2279  \\
J022440.83+400823.0   & & 0221+399    &  &&10.02 &&13621  \\
J023530.74+571524.8   & & 0231+570    &  &&13.68 && 2223  \\
J023619.55+524412.4   & & 0232+525    &  &&13.76 && 6096  \\
J024502.37$-$171220.5 & & 0242$-$174  &  &&15.54 && 7672  \\
J031149.19+190055.7   & & 0308+188    &  &&14.46 && 2711  \\
J031315.18+190824.5   & & 0310+188    &  &&16.20 && 2547  \\
J031445.95+481206.1   & & 0311+480    &  &&14.33 && 4908  \\
J031942.73+344223.8   & & 0316+345    &  &&14.37 && 7868  \\
J034329.01$-$454904.2 & & 0341$-$459  &  &&15.19 &&15506  \\
J035024.96+171447.4   & & 0347+171    &  && 9.47 && 2515  \\
J035630.59$-$364119.7 & & 0354$-$368  &  &&12.66 && 9400  \\
J035705.82+283751.5   & & 0353+284    &  &&11.67 && 5606  \\
J040434.12+250851.8   & & 0401+250    &  &&13.58 && 3841  \\
J041010.32+180223.8   & & 0407+179    &  &&14.50 && 2303  \\
J044321.26+464205.7   & & 0441+467    &  &&12.76 && 4283  \\
J045013.52+174206.1   & & 0447+176    &  &&12.09 && 3155  \\
J045535.93$-$292900.0 & & 0453$-$295  &  &&15.58 && 9872  \\
J045713.22$-$280752.8 & & 0455$-$282  &  &&13.90 && 9869  \\
J045722.55+415556.6   & & 0453+418    &  &&11.98 && 5475  \\
J050003.17$-$362346.4 & & 0458$-$364  &  &&13.33 &&13986  \\
J050355.38$-$285436.0 & & 0501$-$289  &  &&13.58 && 8629  \\
J050530.60+524951.9   & & 0501+527    &  &&11.72 && 3288  \\
J051233.54+165209.6   & & 0509+168    &  &&13.47 && 2931  \\
J051302.56+162246.8   & & 0510+163    &  &&14.15 && 2930  \\
J052906.46+271257.6   & & 0526+271    &  &&15.17 && 9014  \\
J053244.82+261200.7   & & 0529+261    &  &&14.14 && 7321  \\
J053620.20+412955.7   & & 0532+414    &  &&13.46 && 5935  \\
J054748.47+280311.6   & & 0544+280    &  &&13.04 && 5246  \\
J055814.64$-$373426.1 & & 0556$-$375  &  &&14.64 &&10756  \\
J061000.36+281428.4   & & 0606+282    &  &&13.00 && 3538  \\
J061518.70+174341.9   & & 0612+177    &  &&13.37 && 2676  \\
J061934.22+553642.9   & & 0615+556    &  &&13.40 && 3258  \\
J062312.60$-$374127.9 & & 0621$-$376  &  &&12.09 &&11875  \\
J062702.01$-$252249.7 & & 0625$-$253  &  &&12.98 && 9502  \\
J064112.82+474419.8   & & 0637+477    &  &&14.52 && 2673  \\
J064856.08$-$252347.0 & & 0646$-$253  &  &&13.74 && 9658  \\
J071736.26+582420.4   & & 0713+584    &  &&12.03 && 2993  \\
J073427.45+484115.6   & & 0730+487    &  &&14.96 && 5327  \\
J082705.14+284402.6   & & 0824+288    &  &&14.27 && 9142  \\
J084253.04+230025.6   & & 0839+231    &  &&14.45 && 3552  \\
J084644.40+353833.7   & & 0843+358    &  &&14.72 && 7568  \\
J084909.48+342947.8   & & 0846+346    &  &&15.47 && 6769  \\
J085730.45+401613.2   & & 0854+404    &  &&15.16 &&11310  \\
J090148.65+360708.1   & & 0858+363    &  &&14.87 && 9274  \\
J092921.28$-$041005.9 & & 0926$-$039  &  &&14.57 && 1030  \\
J094159.32+065717.1   & & 0939+071    &  &&15.11 && 1621  \\
J094250.60+260100.1   & & 0939+262    &  &&14.88 && 4173  \\
J094846.64+242126.0   & & 0945+245    &  &&14.47 && 4244  \\
J101628.64$-$052032.8 & & 1013$-$050  &  &&13.21 && 1362  \\
J101801.63+072123.9   & & 1015+076    &  &&15.59 && 1179  \\
J102405.90+262103.7   & & 1021+266    &  && 9.33 && 7073  \\
J102459.84+044610.5   & & 1022+050    &  &&14.16 && 3186  \\
J102712.01+322329.8   & & 1024+326    &  &&13.51 && 9985  \\
J102909.80+020553.7   & & 1026+023    &  &&14.05 && 2885  \\
J103936.73+430609.2   & & 1036+433    &  &&11.17 && 4800  \\
J104616.19$-$034033.4 & & 1043$-$034  &  &&14.14 && 1913  \\
J105220.53$-$160804.3 & & 1049$-$158  &  &&14.59 && 7601  \\
J105443.32+270657.2   & & 1052+273    &  &&13.73 && 4875  \\
J105709.94+301336.8   & & 1054+305    &  &&14.69 && 5270  \\
J110432.58+361049.1   & & 1101+364    &  &&14.87 && 5109  \\
J111912.41+022033.1   & & 1116+026    &  &&14.82 && 2222  \\
J111934.60$-$023903.1 & & 1117$-$023  &  &&14.61 && 3512  \\
J112542.87+422358.3   & & 1122+426    &  &&13.25 && 6053  \\
J112619.09+183917.2   & & 1123+189    &  &&14.20 && 4600  \\
J112910.93+380850.1   & & 1126+384    &  &&15.22 && 9645  \\
J112918.04+181645.8   & & 1126+185    &  &&14.10 && 2932  \\
J113227.35+151731.0   & & 1129+155    &  &&14.26 && 2853  \\
J113423.42+314605.9   & & 1131+320    &  &&14.94 &&10834  \\
J113705.10+294758.1   & & 1134+300    &  &&12.64 && 9241  \\
J114359.35+072906.1   & & 1141+077    &  &&14.47 && 2810  \\
J114803.16+183046.6   & & 1145+187    &  &&14.38 && 7930  \\
J115006.09$-$231613.8 & & 1148$-$230  &  &&14.56 &&14440  \\
J115119.30+125359.8   & & 1148+131    &  &&14.15 && 3680  \\
J115154.20+052839.7   & & 1149+057    &  &&15.37 && 5172  \\
J120145.98$-$034540.6 & & 1159$-$034  &  &&15.04 && 2630  \\
J120526.70$-$233312.3 & & 1202$-$232  &  &&12.90 &&14521  \\
J120936.01$-$033307.6 & & 1207$-$032  &  &&13.69 && 2636  \\
J121229.13$-$062206.8 & & 1209$-$060  &  &&13.39 && 2608  \\
J121233.90+134625.0   & & 1210+140    &  &&14.78 && 3027  \\
J121356.28+325631.6   & & 1211+332    &  &&14.93 &&11489  \\
J121410.52$-$171420.2 & & 1211$-$169  &  &&10.15 &&19365  \\
J122747.36$-$081438.0 & & 1225$-$079  &  &&16.06 && 1347  \\
J123515.36+233419.4   & & 1232+238    &  &&13.63 && 8160  \\
J124428.57$-$011858.1 & & 1241$-$010  &  &&13.51 && 3307  \\
J125217.16+154444.2   & & 1249+160    &  &&15.00 &&10250  \\
J125223.56+175651.6   & & 1249+182    &  &&15.43 &&10340  \\
J125514.83+373229.3   & & 1253+378    &  &&15.58 && 8160  \\
J125702.33+220152.9   & & 1254+223    &  &&13.67 &&18254  \\
J131341.59$-$305133.5 & & 1310$-$305  &  &&14.92 &&14136  \\
J131621.95+290556.3   & & 1314+293    &  &&12.77 && 8381  \\
J132115.12+462324.0   & & 1319+466    &  &&14.97 && 8291  \\
J133601.94+482846.7   & & 1333+487    &  &&13.89 && 8310  \\
J133741.51+363903.8   & & 1335+369    &  &&14.51 && 9836  \\
J133913.55+120831.0   & & 1336+123    &  &&14.89 && 3770  \\
J134117.94+342153.6   & & 1339+346    &  &&14.93 && 8492  \\
J134307.26$-$310151.4 & & 1340$-$307  &  &&13.25 &&11366  \\
J135153.93+140945.6   & & 1349+144    &  &&14.77 && 3288  \\
J141026.96+320836.1   & & 1408+323    &  &&14.22 && 7984  \\
J141329.93+213730.0   & & 1411+218    &  &&13.86 && 5943  \\
J142439.16+091714.2   & & 1422+095    &  &&14.90 && 2935  \\
J143545.65$-$163818.1 & & 1432$-$164  &  &&14.58 &&13190  \\
J144814.07+282511.6   & & 1446+286    &  &&14.71 &&16402  \\
J145156.24+422142.9   & & 1450+425    &  &&15.57 && 8167  \\
J151127.61+320417.9   & & 1509+322    &  &&13.10 && 3865  \\
J151714.27+031028.0   & & 1514+033    &  &&13.79 && 4115  \\
J152950.39+085546.3   & & 1527+090    &  &&14.72 && 3552  \\
J154419.46+180643.9   & & 1542+182    &  &&15.08 && 4373  \\
J155501.99+351328.6   & & 1553+353    &  &&14.74 &&12136  \\
J155804.76$-$090807.3 & & 1555$-$089  &  &&13.37 && 3793  \\
J160521.18+430436.6   & & 1603+432    &  &&15.32 && 1353  \\
J160532.09+122542.8   & & 1603+125    &  &&15.91 && 3148  \\
J161053.25+114353.6   & & 1608+118    &  &&14.61 && 3466  \\
J161419.14$-$083326.4 & & 1611$-$084  &  &&13.43 && 3792  \\
J161623.83+265310.7   & & 1614+270    &  &&14.82 &&14663  \\
J161928.99$-$390711.5 & & 1616$-$390  &  &&14.63 &&11944  \\
J162333.83$-$391346.1 & & 1620$-$391  &  &&11.09 &&12748  \\
J163339.30+393053.6   & & 1631+396    &  &&13.88 &&31052  \\
J164539.13+141746.3   & & 1643+143    &  &&15.69 && 3436  \\
J164718.40+322833.0   & & 1645+325    &  &&13.90 &&27543  \\
J170033.62+441024.3   & & 1659+442    &  &&13.27 &&43025  \\
J170530.69+480311.4   & & 1704+481    &  &&13.93 &&20729  \\
J172643.19+583732.0   & & 1725+586    &  &&13.44 &&10838  \\
J175255.81+094751.9   & & 1750+098    &  && 9.53 && 1425  \\
J175332.27+103724.3   & & 1751+106    &  &&14.15 && 4268  \\
J181140.81+282939.5   & & 1809+284    &  &&14.06 && 5172  \\
J182029.78+580441.2   & & 1819+580    &  &&14.23 && 3511  \\
J182337.00+410402.2   & & 1822+410    &  &&14.63 &&18410  \\
J191858.65+384321.8   & & 1917+386    &  &&11.58 && 1700  \\
J194740.52$-$420026.3 & & 1944$-$421  &  &&10.30 &&23397  \\
J195219.66$-$384613.8 & & 1948$-$389  &  &&13.34 &&32011  \\
J200039.25+014341.9   & & 1958+015    &  &&12.48 && 2948  \\
J202706.23+553415.0   & & 2025+554    &  &&12.98 && 6313  \\
J202956.18+391332.3   & & 2028+390    &  &&12.45 && 2902  \\
J203202.39+183139.6   & & 2029+183    &  &&12.20 &&12648  \\
J203454.59$-$273449.2 & & 2031$-$277  &  &&15.28 && 6762  \\
J203838.16$-$332635.0 & & 2035$-$336  &  &&14.25 &&12714  \\
J204808.16+395137.8   & & 2046+396    &  &&14.94 && 2942  \\
J204906.71+372813.2   & & 2047+372    &  &&12.74 && 2989  \\
J210031.30+505118.0   & & 2058+506    &  &&15.93 && 3907  \\
J211244.06+500618.1   & & 2111+498    &  &&12.93 && 3354  \\
J211652.86+241214.9   & & 2114+239    &  &&12.39 && 4421  \\
J211708.29+341227.6   & & 2115+339    &  &&12.33 && 2117  \\
J211717.80+504407.3   & & 2115+505    &  &&11.55 && 3392  \\
J211856.30+541241.4   & & 2117+539    &  &&11.99 && 6914  \\
J212146.78$-$331048.0 & & 2118$-$333  &  &&14.27 && 7625  \\
J212454.89+155903.8   & & 2122+157    &  &&13.80 &&12929  \\
J212458.14+282603.5   & & 2122+282    &  &&14.60 && 2969  \\
J212743.10$-$221148.4 & & 2124$-$224  &  &&14.94 &&11028  \\
J213636.12+220433.5   & & 2134+218    &  &&14.53 &&13813  \\
J213652.94+124719.5   & & 2134+125    &  &&13.35 &&11732  \\
J213846.20+230917.6   & & 2136+229    &  &&12.28 &&14626  \\
J214954.57+281659.8   & & 2147+280    &  &&15.04 &&16920  \\
J215202.73+372617.9   & & 2149+372    &  &&12.59 && 9941  \\
J215453.40$-$302918.4 & & 2151$-$307  &  &&15.05 && 7776  \\
J215618.25+410245.5   & & 2154+408    &  &&14.61 && 3191  \\
J220714.40+072232.3   & & 2204+071    &  &&14.91 && 7675  \\
J221029.22$-$300543.7 & & 2207$-$303  &  &&13.61 &&11058  \\
J222919.42$-$444138.4 & & 2226$-$449  &  &&14.48 &&10279  \\
J223822.75+313418.4   & & 2236+313    &  &&14.75 &&11464  \\
J225848.13+251544.0   & & 2256+249    &  &&12.63 &&13430  \\
J230740.13$-$342753.4 & & 2304$-$347  &  &&14.86 &&10752  \\
J231219.65+260419.7   & & 2309+258    &  &&14.57 && 9279  \\
J232606.58+160019.4   & & 2323+157    &  &&13.63 && 4815  \\
J232715.83+400124.7   & & 2324+397    &  &&15.41 &&21231  \\
J233135.65+410130.6   & & 2329+407    &  &&14.18 &&17476  \\
J233149.93$-$285252.6 & & 2329$-$291  &  &&14.29 &&10859  \\
J233536.58$-$161743.8 & & 2333$-$165  &  &&13.57 && 5660  \\
J234350.87+323247.2   & & 2341+322    &  &&13.28 &&11404  \\
J235530.18$-$251612.7 & & 2352$-$255  &  &&13.61 &&10780  \\
J235644.76$-$301631.6 & & 2354$-$305  &  &&15.01 &&10635  \\ 
\end{supertabular} \\
\vspace{2mm} 
{WDs observed by WASP, including the WASP identity, corresponding
identity in the McCook \& Sion catalogue, WASP magnitude, and the
number of individual data points contributing to the light curve in
the WASP archive. The WASP magnitude is defined
as-2.5log$_{10}(F/10^6)$, where $F$ is the mean WASP flux in
 $\mu$Vega; it is a pseudo-V magnitude comparable to the Tycho-V
 magnitude.}
\end{center}

\twocolumn
\bibliographystyle{mn2e}
\bibliography{Faedi}

\label{lastpage} 
\end{document}